\documentclass[conference]{IEEEtran} 

\usepackage{graphicx}
\usepackage{amsmath, amssymb} 
\usepackage{hyperref} 
\usepackage[utf8]{inputenc} 
\usepackage{booktabs} 
\usepackage{float} 
\usepackage{algorithm}
\usepackage{algpseudocode} 
\usepackage{textcomp} 
\usepackage{balance} 
\usepackage{subcaption} 
\usepackage{xcolor} 
\usepackage{enumitem} 

\graphicspath{{images/}}

\title{DP-RTFL: Differentially Private Resilient Temporal Federated Learning for Trustworthy AI in Regulated Industries}

\author{
    \IEEEauthorblockN{Abhijit Talluri}
    \IEEEauthorblockA{\textit{RTFL Project Contributor} \\
    Email: \texttt{talluri.abhijit@gmail.com}} 
    \IEEEauthorblockA{\textit{\small{Developmental insights reflect work up to March 1, 2024.}}}
}

\begin{document}
\maketitle

\begin{abstract}
Federated Learning (FL) has emerged as a critical paradigm for enabling privacy-preserving machine learning, particularly in regulated sectors such as finance and healthcare. However, standard FL strategies often encounter significant operational challenges related to fault tolerance, system resilience against concurrent client and server failures, and the provision of robust, verifiable privacy guarantees essential for handling sensitive data. These deficiencies can lead to training disruptions, data loss, compromised model integrity, and non-compliance with data protection regulations (e.g., GDPR, CCPA). This paper introduces Differentially Private Resilient Temporal Federated Learning (DP-RTFL), an advanced FL framework designed to ensure training continuity, precise state recovery, and strong data privacy. DP-RTFL integrates local Differential Privacy (LDP) at the client level with resilient temporal state management and integrity verification mechanisms, such as hash-based commitments (referred to as Zero-Knowledge Integrity Proofs or ZKIPs in this context). The framework is particularly suited for critical applications like credit risk assessment using sensitive financial data, aiming to be operationally robust, auditable, and scalable for enterprise AI deployments. The implementation of the DP-RTFL framework is available as open-source \cite{Talluri_DP_RTFL_Framework_2024}.
\end{abstract}


\section{Introduction}
Federated Learning (FL) enables collaborative AI model development across distributed data sources without centralizing sensitive raw data \cite{mcmahan2017communication}. This paradigm is particularly promising for industries like finance and healthcare, which handle highly sensitive information and are subject to strict data privacy regulations. Despite its potential, the widespread adoption of FL in mission-critical sectors is hindered by significant challenges, primarily operational resilience against concurrent client and server failures, and the lack of robust, verifiable privacy guarantees. Standard FL systems may suffer training disruptions, data loss, and model integrity issues, and often struggle to comply with data protection mandates such as GDPR \cite{gdpr} and CCPA.

To address these deficiencies, we propose Differentially Private Resilient Temporal Federated Learning (DP-RTFL). DP-RTFL is a comprehensive framework designed for continuous, auditable, and privacy-preserving training on sensitive datasets, such as financial records for credit risk assessment (e.g., the Credit Card Approval Prediction dataset from Kaggle \cite{kaggleCreditCard}). Our approach enhances the original Resilient Temporal Federated Learning (RTFL) concept by deeply integrating Local Differential Privacy (LDP) at the client level. This ensures that individual contributions to model updates are cryptographically protected before aggregation, complementing the framework's inherent resilience capabilities. The software implementation and experimental code for DP-RTFL are publicly available \cite{Talluri_DP_RTFL_Framework_2024}.

The key contributions of DP-RTFL include:
\begin{itemize}[leftmargin=*]
    \item \textbf{Local Differential Privacy (LDP) Integration:} Client-side application of $(\epsilon, \delta)$-Differential Privacy to model updates, formally limiting information leakage about individual data records and supporting regulatory compliance.
    \item \textbf{Temporal Checkpoint Manifold (TCM):} A distributed, chronologically ordered log of global model states and client contributions, allowing precise rollback for recovery and auditability.
    \item \textbf{Differential State Synchronization (DSS):} Clients transmit only model parameter changes (deltas), reducing communication overhead, which is especially crucial when updates are expanded by DP noise.
    \item \textbf{Adaptive Role Reassignment Protocol (ARRP):} A dynamic protocol enabling eligible clients to assume coordination duties if the central server fails, ensuring training continuity.
    \item \textbf{Zero-Knowledge Integrity Proofs (ZKIP):} A mechanism (in this implementation, a hash-based commitment) accompanying model updates, allowing verification of integrity and origin without revealing sensitive data, thereby enhancing verifiability.
    \item \textbf{Entropy-Based Corruption Detection (EBCD):} Utilizes information-theoretic principles (statistical moments) to identify corrupted or anomalous model updates, even in the presence of DP noise.
\end{itemize}
This paper details the DP-RTFL framework, its components, system architecture, application to credit risk assessment, and a comprehensive evaluation strategy.

\section{Related Work}
Federated Learning, introduced by McMahan et al. \cite{mcmahan2017communication}, laid the groundwork for distributed machine learning with data decentralization. Kairouz et al. \cite{kairouz2019advances} provide an extensive overview of advances and open problems in FL, highlighting challenges in efficiency, robustness, and privacy.

The need for privacy in FL has led to the integration of Differential Privacy (DP) \cite{dwork2006calibrating, dwork2014algorithmic}. Local Differential Privacy (LDP) is particularly relevant as it provides privacy guarantees at the client level before data (or model updates) leave the user's device. Works such as Truex et al. \cite{truex2019hybrid} and Sun et al. \cite{sun2021ldp} have explored LDP in FL, though practical implementations face challenges with utility and high dimensionality. DP-RTFL aims to provide a practical LDP integration by carefully calibrating noise and combining it with other protective measures.

Resilience and fault tolerance are critical for enterprise FL deployments. While some works focus on Byzantine fault tolerance in aggregation \cite{blanchard2017machine, yin2018byzantine}, DP-RTFL's ARRP addresses server/coordinator failures, and its TCM provides a general mechanism for state recovery from various faults. The concept of temporal checkpointing for resilience is also explored in distributed systems, although its specific application as a manifold in FL, as in TCM, is a nuanced contribution.

Integrity verification in FL is crucial. While full Zero-Knowledge Proofs (ZKPs) can offer strong guarantees for verifiable computation in machine learning (ZKML) \cite{ghodsi2017safetynets, liu2021जामा}, they can be computationally intensive. DP-RTFL employs a lighter-weight hash-based commitment scheme as its ZKIP component (\texttt{zkip.py}), akin to those used for data integrity checks \cite{merkle1988digital}, to verify update authenticity and integrity from known participants in a federated setup. Naseri et al. \cite{naseri2023towards} discuss the broader need for robust and verifiable FL.

Anomaly detection using statistical properties, such as EBCD's use of variance, kurtosis, and skewness (\texttt{ebcd.py}), draws from established statistical process control and data mining techniques \cite{chandola2009anomaly, jobson1992applied}. Adapting these for noisy, high-dimensional model parameter distributions in DP-FL is a specific challenge that EBCD addresses.

DP-RTFL distinguishes itself by holistically combining these elements—LDP, temporal resilience, adaptive coordination, hash-based integrity proofs, and information-theoretic anomaly detection—into a unified framework tailored for regulated industries.

\section{The DP-RTFL Framework}
DP-RTFL is designed with several key components working in concert to provide a robust, private, and auditable FL environment. These modules, illustrated conceptually in the overall system architecture (see Figure~\ref{fig:system_architecture_overview}), interact to deliver the framework's capabilities.

\subsection{Local Differential Privacy (LDP)}
LDP is applied client-side to protect individual data contributions before model updates are sent to the aggregator.
\begin{itemize}[leftmargin=*]
    \item \textbf{Mechanism:} Each client, after local training with libraries like scikit-learn for model implementation and NumPy for numerical operations (\texttt{fl\_client.py}), computes model parameter deltas. These deltas are then privatized using an $(\epsilon, \delta)$-Differential Privacy mechanism. This involves L2 norm clipping of the deltas, followed by the addition of Gaussian noise calibrated to the sensitivity of the clipped deltas and the chosen privacy budget $(\epsilon, \delta)$.
    \item \textbf{Purpose:} This ensures that the server or any other entity cannot infer significant information about any individual's data from their (noisy) model update, which is crucial for regulatory compliance and user trust.
    \item \textbf{Configuration:} The privacy budget $(\epsilon, \delta)$ and clipping norm are configurable per client or per round, allowing a trade-off between privacy and model utility.
\end{itemize}
Algorithm~\ref{alg:ldp} outlines the client-side LDP process, as implemented in \texttt{fl\_client.py}.

\begin{algorithm}[htbp]
\caption{Client-Side LDP for Model Deltas}
\label{alg:ldp}
\begin{algorithmic}[1]
\State \textbf{Input:} Local model parameters $\theta_{local}$, base model parameters $\theta_{base}$, L2 clip bound $C$, privacy budget $(\epsilon, \delta)$
\State Compute delta: $\Delta \gets \theta_{local} - \theta_{base}$ (via DSS)
\State Compute L2 norm: $N_2(\Delta) \gets \sqrt{\sum \Delta_i^2}$
\State Clip delta: $\Delta_{clipped} \gets \Delta \cdot \min(1, C / (N_2(\Delta) + \text{1e-6}))$
\State Calculate noise stddev: $\sigma \gets \frac{C \sqrt{2 \ln(1.25 / \delta)}}{\epsilon}$ (if $\epsilon > 0$)
\State Generate Gaussian noise: $Noise \sim \mathcal{N}(0, \sigma^2)$
\State Add noise: $\Delta_{private} \gets \Delta_{clipped} + Noise$
\State \textbf{Output:} Private delta $\Delta_{private}$
\end{algorithmic}
\end{algorithm}

\subsection{Temporal Checkpoint Manifold (TCM)}
The TCM (\texttt{tcm.py}) provides robust, auditable, and precise recovery of global model states and training history. It maintains a chronological log of all global model states, client update summaries (including DP parameters used), and coordinator actions. Each entry is timestamped and hashed for integrity, enabling precise rollback and providing comprehensive audit trails.

\subsection{Differential State Synchronization (DSS)}
DSS (\texttt{dss.py}) aims to reduce communication overhead by transmitting only model parameter deltas. Clients compute the difference (delta) between their updated local model and the global model received. Only this delta is transmitted, minimizing data transfer, especially when updates are expanded by DP noise.

\subsection{Adaptive Role Reassignment Protocol (ARRP)}
ARRP (\texttt{arrp.py}) ensures uninterrupted training by dynamically reassigning the coordinator role if the central server fails. It involves automatic detection of server failure and an election process among eligible, active clients to select a new coordinator, thereby providing seamless failover.

\subsection{Zero-Knowledge Integrity Proofs (ZKIP)}
The ZKIP component (\texttt{zkip.py}) in DP-RTFL uses a hash-based commitment scheme. Each client generates a SHA256 hash of its serialized (noisy) model delta concatenated with a shared secret. This proof accompanies the delta, allowing the coordinator to verify its integrity and origin from a legitimate participant without needing to know the original (pre-noise) delta. This process ensures that updates are authentic and untampered.

\subsection{Entropy-Based Corruption Detection (EBCD)}
EBCD (\texttt{ebcd.py}) detects potentially corrupted model updates by monitoring statistical moments (variance, kurtosis, and skewness) of model parameters. It establishes a dynamic baseline from initial client models and flags deviations that exceed a tolerance factor, helping identify anomalies beyond expected DP noise. The use of such moments for anomaly detection is a common technique in statistical analysis \cite{jobson1992applied, chandola2009anomaly}.

\subsection{Early Stopping}
Standard early stopping (\texttt{earlystop.py}) is employed at client and server levels to prevent overfitting and improve efficiency. It halts training or restores the best model if validation performance (loss for clients, accuracy for server) ceases to improve for a configurable patience period.

\section{Threat Model and Assumptions}
DP-RTFL operates under the following threat model:
\begin{itemize}[leftmargin=*]
    \item \textbf{Clients:} Assumed to be honest-but-curious regarding their own data. However, a fraction of clients could be malicious (Byzantine), attempting to degrade model performance or integrity. LDP provides protection against inference from their updates regardless of server behavior. EBCD and ZKIPs aim to mitigate impacts from overtly malicious updates.
    \item \textbf{Server/Coordinator:} The central server or an ARRP-elected coordinator is assumed to be honest-but-curious. It will follow the protocol but might attempt to infer information about individual client data from received updates. LDP is the primary defense against this. The server is also a point of failure, addressed by ARRP and TCM.
    \item \textbf{External Adversaries:} May attempt to eavesdrop on communication channels or compromise clients/server. Secure communication channels (e.g., TLS, not explicitly part of this framework's core logic but assumed in a real deployment) would be necessary. ZKIPs help ensure update integrity against tampering in transit.
\end{itemize}
It is assumed that cryptographic primitives (hashing for ZKIP) are secure and that the shared secret for ZKIP is managed securely among legitimate participants.

\section{System Architecture}
\begin{figure*}[htbp] 
    \centering
    \includegraphics[width=0.9\textwidth]{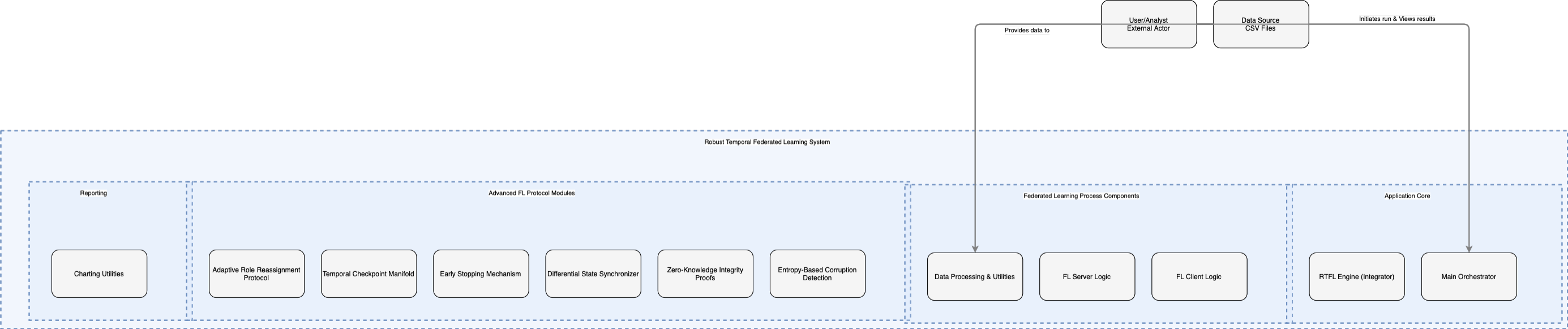} 
    \caption{DP-RTFL Conceptual System Diagram. This diagram visually represents the architecture detailed in the project's \texttt{README.md}, showing data flow and component interactions (adapted from the project's \texttt{README.md}).}
    \label{fig:system_architecture_overview}
\end{figure*}

The DP-RTFL system architecture is designed for modularity and resilience. A conceptual overview, illustrating the interaction between data sources, core FL processes, advanced protocol modules, and reporting utilities, is presented in Figure~\ref{fig:system_architecture_overview}. This diagram visually represents the system components detailed in the project's \texttt{README.md} file.

The main components are:
\begin{enumerate}[leftmargin=*,label=\arabic*)]
    \item \textbf{Data Sources:} Distributed datasets, e.g., \texttt{application\_record.csv} and \texttt{credit\_record.csv} for credit risk from a source like Kaggle \cite{kaggleCreditCard}.
    \item \textbf{Application Core (\texttt{main.py}):} Orchestrates the simulation, client/server instances, metrics collection, and integrates all DP-RTFL components.
    \item \textbf{Federated Learning Process Components:}
        \begin{itemize}[leftmargin=*]
            \item \textit{FL Server Logic (\texttt{fl\_server.py}):} Manages aggregation, ZKIP verification, server-side EBCD, TCM, ARRP, and global evaluation.
            \item \textit{FL Client Logic (\texttt{fl\_client.py}):} Handles local training, LDP, ZKIP generation, and DSS interaction.
            \item \textit{Data Utilities (\texttt{data\_utils.py}):} Data loading, preprocessing, and client splitting.
        \end{itemize}
    \item \textbf{Advanced FL Protocol Modules:} TCM, DSS, ARRP, ZKIP, EBCD, and Early Stopping, each implemented in their respective Python files (e.g., \texttt{tcm.py}, \texttt{dss.py}).
    \item \textbf{Reporting (\texttt{charting.py}):} Generates plots for metrics visualization.
\end{enumerate}

\section{Methodology and Experimental Setup}
The DP-RTFL methodology encompasses data handling, the federated training process, and simulation parameters.

\subsection{Data Preparation and Distribution}
The primary dataset used is the "Credit Card Approval Prediction" dataset from Kaggle \cite{kaggleCreditCard}, which includes \texttt{application\_record.csv} and \texttt{credit\_record.csv}. Preprocessing steps involve merging these files; deriving a binary target variable for credit risk (good/bad); handling missing values (median imputation for numerical, mode for categorical); and transforming features (e.g., \texttt{DAYS\_BIRTH} to years, \texttt{DAYS\_EMPLOYED} to years of employment, and normalizing negative day counts). Categorical features are one-hot encoded, and numerical features are standardized using scikit-learn's \texttt{ColumnTransformer} and \texttt{StandardScaler}. The processed data is split into a global training set and a test set (80/20 split). The global training set is then distributed (potentially non-IID) among \texttt{NUM\_CLIENTS}. Clients may further split their local data for validation.

\subsection{Federated Training and Simulation}
The simulation (\texttt{main.py}) runs for \texttt{NUM\_ROUNDS}. In each round:
\begin{enumerate}[leftmargin=*]
    \item The server (or ARRP coordinator) provides global parameters. TCM is used for recovery if the server fails to provide parameters.
    \item Active clients (simulating potential dropouts) train a local \texttt{SGDClassifier} model (from scikit-learn) for \texttt{CLIENT\_EPOCHS}.
    \item Deltas are computed (DSS), privatized (LDP with \texttt{DP\_EPSILON}, \texttt{DP\_DELTA}, \texttt{DP\_L2\_NORM\_CLIP}), and a ZKIP is generated.
    \item The server verifies ZKIPs, aggregates valid deltas (weighted by client data sizes), and updates the global model.
    \item EBCD checks the global model, and TCM logs the state.
    \item Metrics (accuracy, F1-score, AUC, etc.) are collected.
\end{enumerate}
Key parameters, such as \texttt{NUM\_CLIENTS=5}, \texttt{NUM\_ROUNDS=10}, \texttt{CLIENT\_EPOCHS=3}, and DP parameters, are specified in \texttt{main.py}.

\section{Evaluation Metrics and Expected Results}
The DP-RTFL framework is evaluated on several axes.

\subsection{Privacy-Utility Trade-off}
\begin{figure}[htbp]
    \centering
    \includegraphics[width=\linewidth]{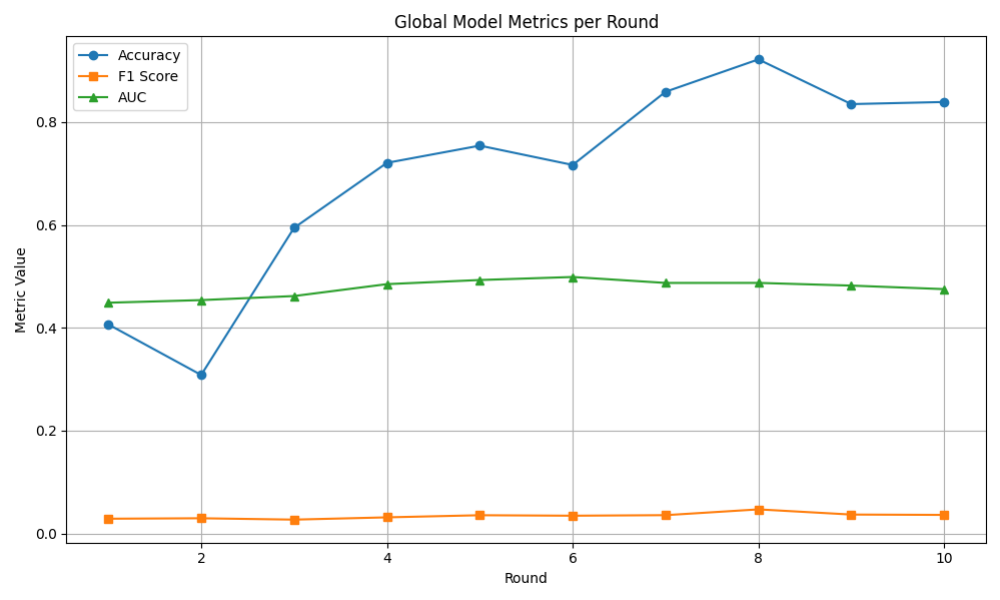}
    \caption{Global Model Metrics (Accuracy, F1, AUC) per Round. Expected to show convergence and impact of DP budget on final performance.}
    \label{fig:eval_global_metrics}
\end{figure}

\begin{figure}[htbp]
    \centering
    \includegraphics[width=0.9\linewidth]{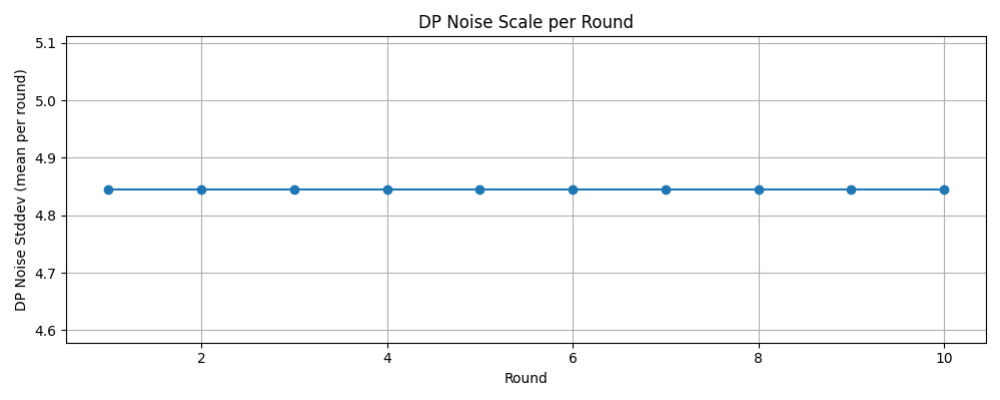}
    \caption{DP Noise Scale (Mean Stddev) per Round. Expected to correlate with chosen privacy parameters ($\epsilon, \delta$).}
    \label{fig:eval_dp_noise_scale}
\end{figure}

This is assessed by model performance (Accuracy, F1-score, AUC-ROC on the test set) against varying DP budgets ($\epsilon, \delta$) and L2 clipping norms.
\begin{itemize}[leftmargin=*]
    \item \textbf{Expected Outcome:} Figure~\ref{fig:eval_global_metrics} would show model convergence. Higher privacy (lower $\epsilon$) is expected to result in a graceful degradation of utility. Figure~\ref{fig:eval_dp_noise_scale}, visualizing mean DP noise standard deviation per round, should show higher noise for stricter privacy budgets.
\end{itemize}

\subsection{Resilience and Recovery Precision}
\begin{figure}[htbp]
    \centering
    \includegraphics[width=0.9\linewidth]{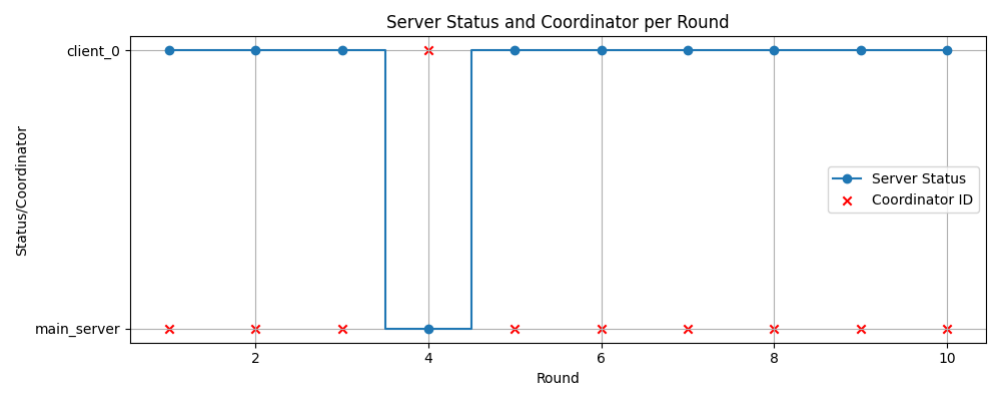}
    \caption{Server Status and Coordinator ID per Round. Expected to show ARRP's dynamic role reassignment during simulated failures.}
    \label{fig:eval_server_status}
\end{figure}

\begin{figure}[htbp]
    \centering
    \includegraphics[width=0.9\linewidth]{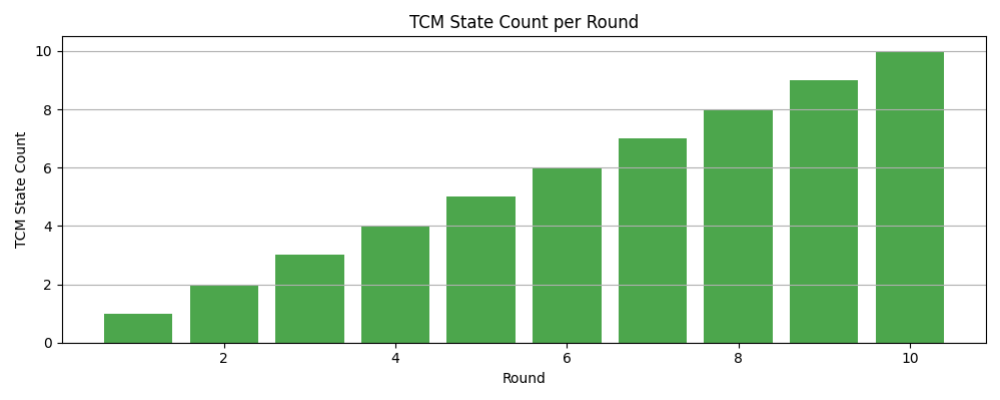}
    \caption{TCM State Count per Round. Expected to show steady accumulation of states, vital for audit and recovery.}
    \label{fig:eval_tcm_state_count}
\end{figure}

Training continuity under simulated failures and TCM recovery fidelity are key metrics.
\begin{itemize}[leftmargin=*]
    \item \textbf{Expected Outcome:} Figure~\ref{fig:eval_server_status} (plotting server status and coordinator ID) should demonstrate ARRP's ability to maintain a coordinator. Figure~\ref{fig:eval_tcm_state_count} (TCM states per round) should show consistent logging. Successful TCM recovery from a mid-training round, as simulated in \texttt{main.py}, would validate recovery precision.
\end{itemize}

\subsection{Communication, Integrity, and Anomaly Detection}
\begin{figure}[htbp]
    \centering
    \includegraphics[width=0.9\linewidth]{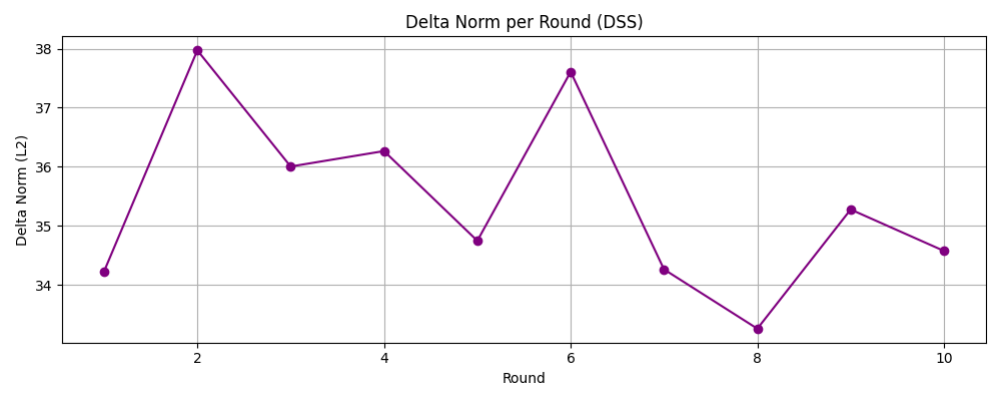}
    \caption{Aggregated Delta Norm (L2) per Round. Reflects the magnitude of global model updates via DSS.}
    \label{fig:eval_delta_norm}
\end{figure}

\begin{figure}[htbp]
    \centering
    \includegraphics[width=0.9\linewidth]{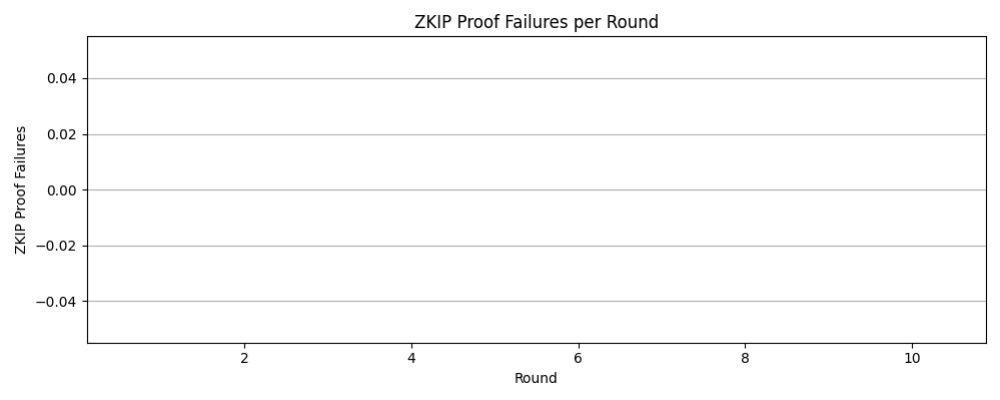}
    \caption{ZKIP Proof Failures per Round. Expected to be low, indicating update integrity.}
    \label{fig:eval_zkip_failures}
\end{figure}

\begin{figure}[htbp]
    \centering
    \includegraphics[width=\linewidth]{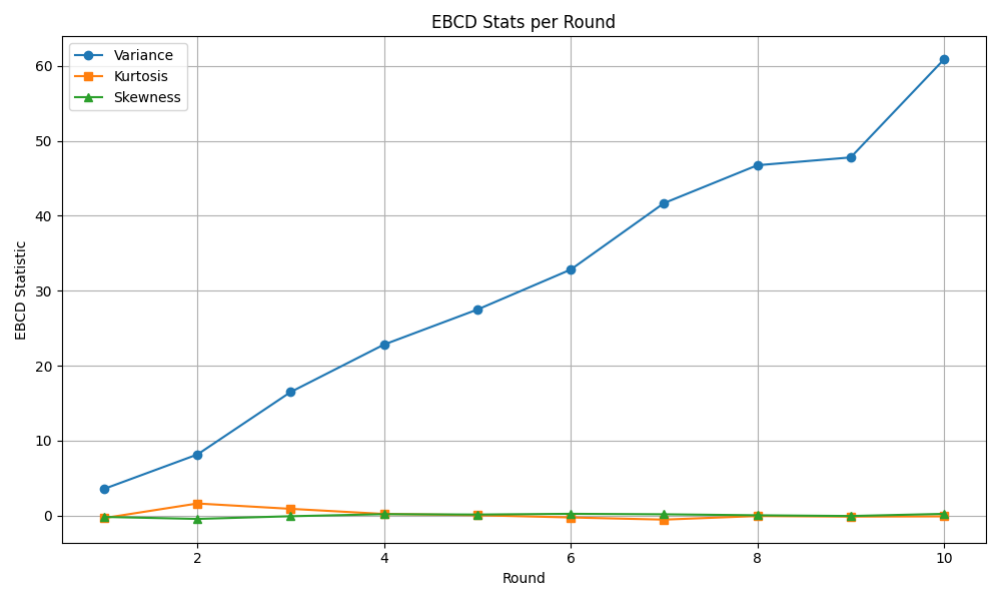}
    \caption{EBCD Statistics (Variance, Kurtosis, Skewness) per Round. Monitors global model parameter distribution.}
    \label{fig:eval_ebcd_stats}
\end{figure}

\begin{figure}[htbp]
    \centering
    \includegraphics[width=0.9\linewidth]{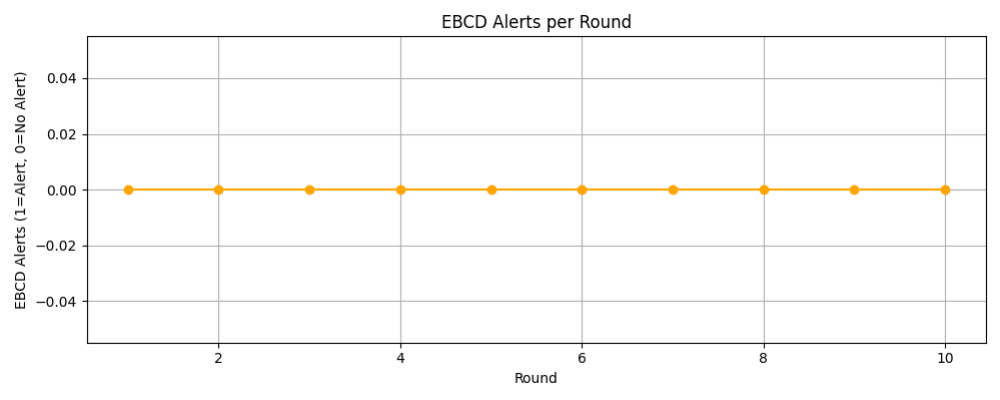}
    \caption{EBCD Alerts per Round. Indicates rounds where potential anomalies were flagged.}
    \label{fig:eval_ebcd_alerts}
\end{figure}

The overhead of DSS, ZKIP, and EBCD effectiveness are measured.
\begin{itemize}[leftmargin=*]
    \item \textbf{Expected Outcome:} Figure~\ref{fig:eval_delta_norm} (L2 norm of aggregated deltas) would show update magnitudes. Figure~\ref{fig:eval_zkip_failures} (ZKIP failures per round) should be minimal in normal operation. Figure~\ref{fig:eval_ebcd_stats} (variance, kurtosis, and skewness of global weights) and Fig.~\ref{fig:eval_ebcd_alerts} (EBCD alerts) should indicate stability and detection of simulated anomalies.
\end{itemize}

\subsection{Training Stability and Auditability}
\begin{figure}[htbp]
    \centering
    \includegraphics[width=0.9\linewidth]{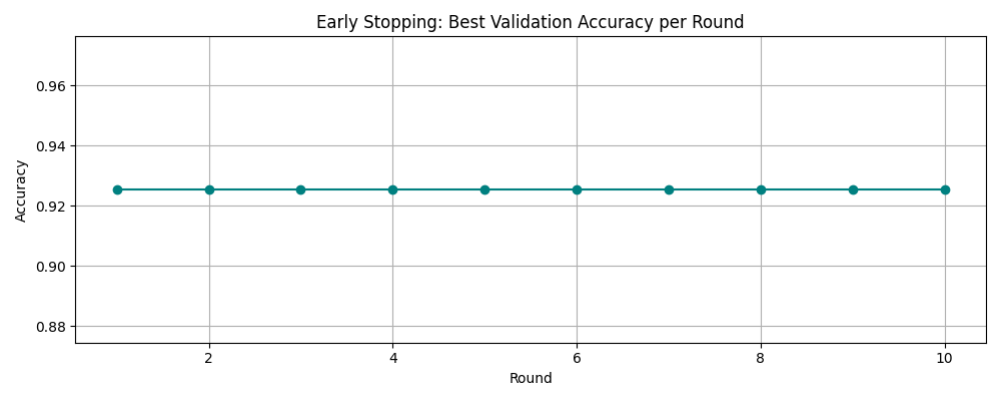}
    \caption{Early Stopping: Best Server Validation Accuracy per Round. Shows effectiveness in finding optimal training duration.}
    \label{fig:eval_early_stopping}
\end{figure}

Early stopping effectiveness and TCM's audit trail completeness are assessed.
\begin{itemize}[leftmargin=*]
    \item \textbf{Expected Outcome:} Figure~\ref{fig:eval_early_stopping} (best validation accuracy from server's early stopping) should demonstrate prevention of overfitting. The TCM logs themselves (content not plotted but functionality tested by recovery) provide the basis for auditability.
\end{itemize}
Per-client plots (e.g., update norms, EBCD stats, ZKIP status from \texttt{charting.py}) would offer granular insights into individual client contributions and adherence to protocols.

\section{Limitations}
While DP-RTFL offers a comprehensive approach, certain limitations exist:
\begin{itemize}[leftmargin=*]
    \item The ZKIP mechanism implemented is a hash-based commitment, which relies on a shared secret and primarily ensures integrity and authenticity from known participants, rather than providing full zero-knowledge computational proofs against arbitrary verifiers.
    \item The framework's evaluation is currently simulation-based. Real-world network conditions, diverse hardware, and true adversarial behaviors might present additional challenges.
    \item The scalability of some components (e.g., TCM logging, EBCD baseline establishment with many clients) needs further testing in very large-scale federations.
    \item The choice of DP parameters ($\epsilon, \delta$, clipping bound) involves a trade-off that might require careful tuning for specific applications and datasets to achieve optimal privacy and utility.
\end{itemize}

\section{Ethical Considerations}
The development and deployment of FL systems, especially in sensitive areas like finance, carry significant ethical responsibilities.
\begin{itemize}[leftmargin=*]
    \item \textbf{Privacy Preservation:} LDP is a core component aiming to protect individual financial data. However, the choice of privacy parameters and the potential for re-identification in high-dimensional sparse data, even with DP, must be continually assessed.
    \item \textbf{Fairness and Bias:} FL models can inherit or even exacerbate biases present in distributed client data. While not a direct focus of DP-RTFL's current components, ensuring fairness and mitigating bias in credit risk assessment models is a critical ongoing concern that necessitates additional mechanisms.
    \item \textbf{Transparency and Auditability:} TCM aims to provide audit trails, contributing to transparency. Nevertheless, the complexity of FL and DP can make full transparency to end-users challenging.
    \item \textbf{Security:} While resilience and integrity are addressed, sophisticated attacks against FL systems represent an evolving research area.
\end{itemize}
Continuous vigilance and adherence to ethical AI principles are necessary throughout the lifecycle of such systems.

\section{Conclusion}
DP-RTFL presents a holistic framework for conducting Federated Learning in environments with stringent requirements for privacy, resilience, and verifiability. By integrating Local Differential Privacy, Temporal Checkpoint Manifolds, Differential State Synchronization, Adaptive Role Reassignment, hash-based integrity proofs (ZKIP), and Entropy-Based Corruption Detection, DP-RTFL addresses critical operational challenges that have historically hindered FL adoption in sensitive domains like finance and healthcare.

The framework's design emphasizes training continuity, precise state recovery, and formal data privacy guarantees, rendering it suitable for applications such as credit risk assessment using sensitive financial data. The potential scientific impact lies in advancing trustworthy distributed AI. Practically, DP-RTFL aims to enable FL adoption for high-stakes applications by satisfying regulatory compliance and reducing operational risks.

Future work will focus on extending LDP schemes and exploring advanced privacy accounting. Investigation into more sophisticated, yet practical, ZKP constructions is warranted. Extensive testing in enterprise-grade environments with diverse datasets is crucial. Exploring fairness-aware mechanisms within the DP-RTFL framework and adapting components for emerging hardware, such as confidential computing enclaves, are other promising directions. Addressing the identified limitations, particularly concerning real-world deployment complexities and advanced adversarial scenarios, will be key to maturing the framework.

\bibliographystyle{IEEEtran}
\bibliography{references}
\balance

\end{document}